# Audio-only Bird Species Automated Identification Method with Limited Training Data Based on Multi-Channel Deep Convolutional Neural Networks

XIE Jiang-jian[1] DING Chang-qing[2]; LI Wen-bin[1*]; CAI Cheng-hao[1]

(1 School of Technology，Beijing Forestry University, Beijing, 100083, P. R. China. 2 School of Nature Conservation，Beijing Forestry University, Beijing, 100083, P. R. China.)

**Abstract**

1. Deep convolutional neural networks (DCNN) have achieved breakthrough performance on bird species identification tasks based on spectrogram features, but a huge number of labeled samples are needed to train an excellent DCNN model. However, it is difficult to collect enough samples for certain bird species. For practical uses of bird species identification, it is significant to study a method which only requires a small sample set.

2. Transfer learning is an available solution to train deep learning models with limited samples. Based on the parameter transfer mode, we design a bird species identification model that uses the VGG-16 model (pretrained on ImageNet) for feature extraction, then a classifier consisting of two fully-connected hidden layers and a Softmax layer is attached. We take the vocalization signals of eighteen bird species which were recorded in Beijing Song-Shan National Nature Reserve as an example, comparing the performance of the transfer learning model with the original VGG16 model. The results show that the former has higher train efficiency, but lower mean average precisions (MAP).

3．To improve the MAP of the transfer learning model, we investigate two fusion modes to form multi-channel identification models. Then we evaluate the models on our own sample sets. We find that the result fusion mode outperforms the feature fusion mode, and the best MAP reaches 0.9998. The number of model parameters is 13110, which is only 0.0082% of the VGG16 model. Also, the size demand of sample is decreased.

4. The type and duration of spectrogram may affect the performance of identification. We choose three kinds of time frequency transformation methods, including Short Time Fourier Transform, Mel-frequency Cepstrum Transform and Chirplet Transform, to calculate the spectrogram. Then the spectrogram is segmented to the duration of 100ms, 300ms and 500ms. Chirplet spectrogram improves training efficiency and the MAP, which is more suitable for the feature representation of bird vocalization. We choose three durations, 100ms, 300ms and 500ms for comparison, the result reveals that the 300ms duration is the best. The duration should be determined based on the time domain characteristic of bird vocalization.

**Tweetable Abstract:**

Based on the transfer learning, we design a bird species identification model that uses the VGG-16 model (pretrained on ImageNet) for feature extraction, then a classifier consisting of two fully-connected hidden layers and a Softmax layer is attached. We compare the performance of the proposed model with the original VGG16 model. The results show that the former has higher train efficiency, but lower mean average precisions(MAP). To improve the MAP of the proposed model, we investigate the result fusion mode to form multi-channel identification model, the best MAP reaches 0.9998. The number of model parameters is 13110, which is only 0.0082% of the VGG16 model. Also, the size demand of sample is decreased.

# Intorduction

Birds have been widely regarded as important indicators of biodiversity because they provide vital ecosystem services. They are susceptible to environmental changes, and these changes are relatively easy to be observed(Priyadarshani, Marsland, & Castro, 2018). The composition, quantity and diversity of bird species can directly reflect habitat suitability, ecosystem health and biodiversity and the quality of regional ecological environments. Therefore, it is significant to investigate and monitor bird species. At the species level, bird vocalizations are relatively stable. There exists distinct diversity of species identification(Green & Marler, 1961) . Also, automated bird vocalization classification has been demonstrated to be useful for species identification(Graciarena, Delplanche, Shriberg, & Stolcke, 2011).

As the autonomous recording units (ARUs) become more popular, passive acoustic monitoring (PAM) has become a powerful method to monitor and survey the birds. PAM is a non-invasive, large spatial and temporal scales method using remote ARUs, which has a very good prospect for application(Kalan et al., 2015). Facing the situation of a deluge of audio data from long-term recording programs, the monitoring task becomes not manageable without an efficient automated identification method.

The key point of automated bird species identification is the extraction of identifiable features of bird vocalizations. Deep learning has a strong self-learning and signal extraction ability, and it can automatically acquire and combine characteristic information from inputs(Chen et al., 2017). Koops et al.(Koops, Van Balen, Wiering, & Multimedia, 2014)released three datasets consisting of Mel-Frequency Cepstral Coefficients (MFCC) and their combinatorial features, such as the mean and the variance. They were used to train Deep Neural Networks (DNN) with several topologies. It was found that the best network can classify 73% of the segments. Piczak(Piczak, 2016) studied three different Deep Convolutional Neural Networks(DCNN) and a simple ensemble model to complete the LifeCLEF 2016 bird identification task, with the ensemble submission achieving a mean average precision(MAP) of 41.2% and 52.9% for foreground species. Tóth and Czeba(Tóth & Czeba, 2016) fed the spectrograms into a convolutional neural network to realize the classification of bird species. The solution reached a MAP score of over 40% for main species and reached a MAP score of over 33% for main species mixed with background species. Sprengel et al.(Sprengel, Jaggi, Yannic Kilcher, & Hofmann, 2016) processed the background noise by image process methods Then they trained a convolutional neural network to classify the bird species. The MAP score of 0.686 was achieved when identifying the main species of each sound file. When background species were considered as additional prediction targets, the MAP score is decreased to 0.555. All of the above studies showed that the bird identification method based on deep learning is effective. By designing a reasonable network architecture and selecting appropriate input features, its identification accuracy can surpass other classification methods.

In a real environment, the sample size of bird vocalization maybe limited. Especially, for rare bird species, it is more difficult to record their vocalizations. Also, there are regional differences among the vocalization of the birds in different places. Therefore, we should not simply download their vocalizations from websites as supplement data. Compared to the demand of a deep neural network model, the sample size of bird vocalization is relatively small, which tends to cause the overfitting problem when training the neural network. In other words, the accuracy of the model is high on the training set, but its generalization performance is poor.

Transfer learning has been widely applied to many areas. It transfers existing knowledge from a source field to a target field and uses it to solve learning problems in the target field, with only a few labels or even no labels (Uricchio, Ballan, Seidenari, & Bimbo, 2017; Zhuang, Ping, Qing, & Shi, 2015; Zou, Li, Chen, & Du, 2016). Transfer learning extracts features from a pretrained model, which decreases the number of parameters significantly and reduces the demand for the number of samples. Thus, it can avoid overfitting.

In this paper, we transferred image recognition to bird identification and utilized the spectrograms of bird vocalization as inputs to train identification models. We took the vocalization signals of eighteen bird species which were recorded in Beijing Song-Shan National Nature Reserve as source data, and we set up three sample sets by using three kinds of time-frequency transform methods. Then the transfer learning model based on VGG16 was proposed. Further, multi-channel models based on the transfer learning models were investigated. With three kinds of sample sets, the performance of the proposed models is evaluated. In addition, the effects of different durations of spectrogram were researched.

## 1 Sample set of spectrograms

## 1.1 Details of vocalization signals

In the breeding season, we had recorded the vocalization of wild birds at Beijing Song-Shan National Nature Reserve (east longitude 115°43'44" - 115°50'22", north latitude 40°29'9" - 40°33'35") with digital solid-state recorder Marantz PMD-671 (MARANTZ, Japan) and directional microphone Sennheiser MKH416-P48 (SENNHEISER ELECTRONIC, German) for many years. The vocalization signals are in 16-bit linear WAV format with 44.1kHz sampling rate. In this paper, we selected the vocalization signals of eighteen bird species which are clearly identified by ornithologists. A signal only contains the vocalization of exactly one species, and there is no overlap between vocalizations. Table1 lists the detailed information of eighteen bird species (sorted by the taxonomic order). The column of time means the cumulative time of the vocalization signals.

Tab.1 Vocalization signal details of 18 kinds of bird

| Order | Family | Species | Time(s) |
|---|---|---|---|
| Galloformes | Phasianidae | *Phasianus colchicus* | 12 |
| Cuculiformes | Cuculidae | *C. micropterus* | 13 |
| | | *C. saturatu*) | 52 |
| | | *Cuculus sparverioides* | 34 |
| Passeriformes | Corvidae | *Corvus macrorhynchos* | 27 |
| | | *Urocissa erythroryncha* | 96 |
| | Turdidae | *Phoenicurus auroreus* | 37 |
| | Muscicapidae | *Ficedula zanthopygia* | 61 |
| | | *F. narcissina* | 82 |
| | | *F. elisae* | 49 |
| | Paridae | *P.major* | 54 |
| | | *Parus palustris* | 33 |
| | | *P. montanus* | 38 |
| | | *P. venustulus* | 26 |
| | Sittidae | *S. villosa* | 29 |
| | | *Sitta europaea* | 36 |
| | Emberizidae | *Emberiza godlewskii* | 23 |
| | | *E. elegans* | 71 |

## 1.2 Signal preprocessing

First, the vocalization signal is filtered via a pre-emphasis filter. Its formula is:

$$\hat{x}(n) = x(n) - \lambda \cdot x(n-1) \tag{1}$$

where $\hat{x}(n)$ and $x(n)$ are the n$^{th}$ sample values of a signal before and after pre-emphasis, and $\lambda$ is the preemphasis coefficient (set to 0.95). The pre-emphasis filter is performed to emphasize the higher frequency portion by giving a greater weight to higher frequencies.

The vocalization is then broken into frames and windowed using the Hamming window function. We chose the frame length of 50ms to make sure that at least one fundamental frequency peak is included, and 30% overlap is chosen to divide the vocalization signal into windowed frames.

The primary element of bird vocalization is 'notes' that can be combined into 'syllables' which in turn can constitute 'song types'. Methods for acoustic classification of bird species can broadly be grouped into those that classify individual syllables or song types. Segmenting vocalization into distinct syllables is a crucial step. We performed the segmentation operation in the time domain based on energy, through computing the energy of the signal in each frame. Then the frames with high energy are considered to be syllables.

## 1.3 spectrogram calculation

Discreminal characters of the vocalization signal can be achieved by extracting suitable parameters to form a feature vector. The option of powerful features is the key point for any automatic identification system. Acoustic signals are usually trsnsformed to spectrograms before processing. A spectrogram is a visual mode of representing the signal intensity of a signal over time at different frequencies. The horizontal axis and vertical axis of spectrogram are time and frequency, and the picture shows how the signal develops and changes over time and frequency with color or gray scale.

The spectrogram can be used to characterize the time-frequency characteristics of bird vocalization, which contains more abundant information compared to the independent characteristic of time or frequency domain. Bird vocalizations can be regarded as the special objects in spectrogram, where the characteristic of special objects represents the time-frequency characteristics of bird vocalization. Then bird species identification can be equalized to object recognition in spectrogram.

To compare the performance of an identification model with different spectrograms, three time-frequency transform methods, Short Time Fourier Transform(STFT), Mel-frequency Cepstral Transform (MFCT) and Chirplet Transform (CT)) were utilized to calculate the spectrograms.

STFT is one of the earliest time frequency analysis methods in acoustic signal process, which presents the energy distribution across linear range of frequencies. Based on STFT, MFCT was proposed to represent the approximately logarithmic frequency sensitivity of human hearing(Simonyan & Zisserman, 2014). We can use MFCT to get Mel spectrograms. The MFCCs are 32-dimensional, and the last 31 dimensions are composed to Mel spectrograms. CT is a kind of linear time-frequency representation, which refers to other time frequency representations to assist each atom on the modulated time frequency plane. It is a broad class of filters which includes Wavelets and Fourier basis as particular cases, and there is an obvious advantage in the representation of short-time stationary signal(Bultan, 2002). We carried out the CTs on each frame with fast Chirplet decomposition algorithm(Glotin, Ricard, & Balestriero, 2016), and the calculated wavelet coefficients were used to compose the Chirplet spectrogram.

**1.4 Create sample sets**

We utilized the above three time-frequency transforms to calculate the spectrogram. Figure 1 represents the signal and its spectrograms of *Phoenicurus auroreus*. From up to down, they are a time domain waveform, an STFT spectrogram, a Mel spectrogram and a Chirplet spectrogram.

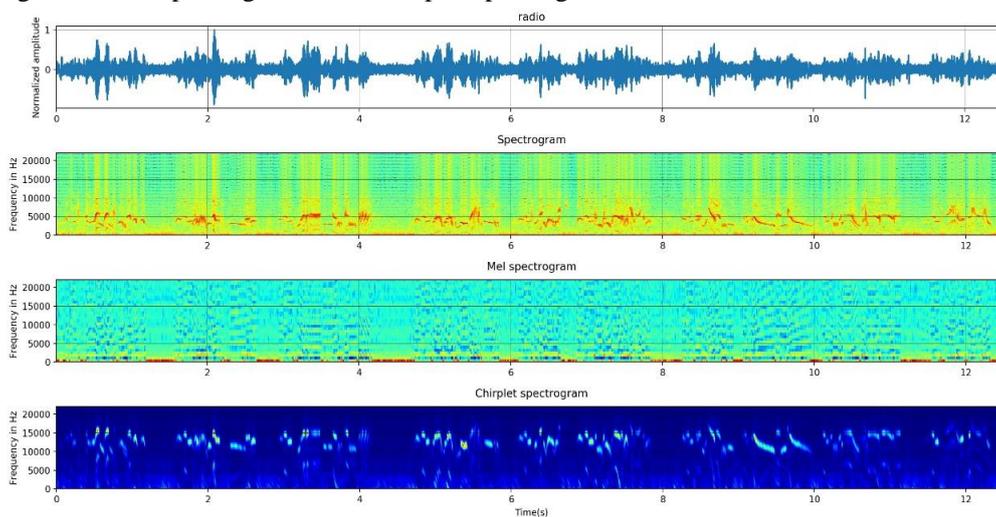

Fig.1 Spectrogram of *Phoenicurus auroreus*

Bird species identification is always regarded as the classification of individual syllables or song types. Potamitis et al.(Potamitis, Ntalampiras, Jahn, & Riede, 2014). Hence, the spectrograms of certain duration are saved as 224 * 224 color images, which are fed into the identification model. Later we will discuss the influence of different durations to the performance of identification models. With three kinds of time-frequency transformers, three sample sets with the same size can be built.

# 2 Bird Sepcies Identification Models

## 2.1 The identification model based on parameter-based transfer learning

As the depth of a deep learning models increases, the number of parameters increases dramatically, so that a huge number of labeled samples are needed during the training. If there are not enough labeled samples, the training will lead to overfitting, and the identification model will be ineffective. In parameter-based transfer learning, a pretrained deep learning model is used for feature extraction. Then a classifier is trained to classify the extracted features. During training, the parameters of featureextraction are frozen, only the parameters of the classifier are updated. The above process reduces significantly the number of trained parameters, so that the size demand of labeled samples decreases. Also, the efficiency of training is improved.

DCNN self-learns the image features via some convolutional and pooling layers. Then it can classify the features by a fully-connected layer and realize the recognition of images. Its local connections, the weight sharing operation and the pooling operation can effectively reduce the complexity of the network and reduce the number of trained parameters(Zhou, Jin, & Dong, 2017). VGG16 is a DCNN model which yielded high performance in the 2014 ImageNet Large Scale Visual Recognition Challenge (ILSVRC)(Simonyan & Zisserman, 2014). It has been widely used in the field of image recognition(Hou, Chen, & Shah, 2017; Triantafyllidou, Nousi, & Tefas, 2017). In this paper, we equalized bird species identification as recognition of vocalization spectrogram, extracting the features of spectrogram with the Imagnet pretrained VGG16 model and feeding the features to two fully-connected layers and a Softmax layer to realize the identification of eighteen bird species. The structure of the proposed model is shown by Figure 2.

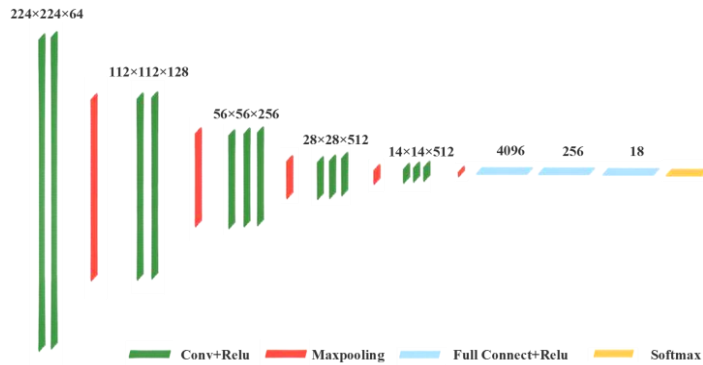

Fig.2 The structure of identification model based on transfer learning

## 2.2 Multi-channel identification models

We further fused three transfer learning models together to improve the efficiency and accuracy of identification. Two fusion modes are shown by Figure 3. The first is feature fusion, which fuses directly the feature outputs of the VGG16 model. The second is result fuse, which fuses Softmax outputs of a transfer learning model. In this method, the parameters of transfer learning are trained independently with diverse kinds of spectrogram. Two kinds of fusion were realized by using two fully-connected layers and a Softmax layer.

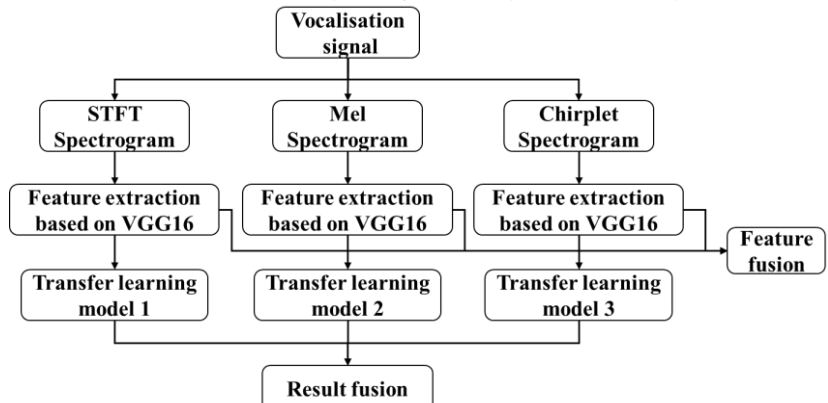

Fig.3 The structure of multi-channel identification models

## 3 Experiments and results

**3.1 Experimental setup**

The experiments were conducted on an Ubuntu16.04 Linux workstation with 32G memory, one E5-2620CPU (6*2.1GHz) and two GTX1080ti GPUs (11GB memory). The model implementation is based on the deep learning framework Tensorflow1.4.1.

The three sample sets are randomly split into training set, validation set and test set with a ratio of 8:1:1. Based on these samples, the identification model is trained and verified. The flowchart of training is shown in Figure 4.

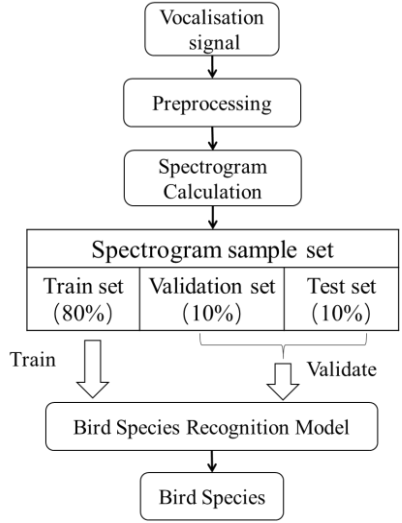

Fig.4 The flowchart of the training of identification models

The training set is divided to several batches to speed up the training process. Considering the memory of the workstation, the size of batch is set to 50. Random initialization is used to initialize the model parameters. The Adam optimization algorithm (Kingma & Ba, 2014). is used to train the parameters, and the initial learning rate is 0.001. The number of training epochs is set to 100.

Because of the different spans of the collected vocalization signals, the number of spectrograms of different bird species in the sample set is quite different. In other words, it is an unbalance sample set which is not beneficial to the training of DCNN models. We introduced the weighted cross-entropy as the loss function of the model. The loss function can increase the weights of the bird species who has few samples, so that the problem of unbalance data can be solved(Cai, Fan, Feris, & Vasconcelos, 2016). The weighted cross-entropy loss function is:

$$\text{WCE} = -\omega y \log \hat{y} - (1-y)\log(1-\hat{y}) \qquad (2)$$

where $\omega$ is the weight vector of labels, and $y$ and $\hat{y}$ are the real label and the predicted label.

**3.2 Different models with the same duration of spectrogram**

We used the VGG16 model as baseline to evaluate the proposed bird species identification method. In the experiments, there are four models, including VGG16, a transfer learning model based on VGG16 (TF), a multi-channel model with result fusion (Re-fuse) and a multi-channel model with feature fusion (Fe-fuse). The spectrogram sample sets of 300ms duration were used to train all the models. Firstly, three spectrogram sample sets, including Ch, Mel and Spe, were fed to the VGG16 model respectively. Then three VGG16 models with different parameters were trained. The same training process is also applied to the TF model.

Figure 5 and Figure 6 provide the variations of validation MAP on different sample sets with VGG16 and TF. The VGG16-Ch model obtains the best MAP (0.9995) during the 51th epoch. On the other hands, the VGG16-Mel model obtains its best MAP (0.9616) during the 81th epoch, and the VGG16-Spe model obtains its best MAP (0.9435) during the 92th epoch. In comparison with the above two models, the VGG16-Ch model has a higher MAP, and its training process are more effective. It demonstrates that the Ch spectrogram sample set is more suitable for bird species identification.

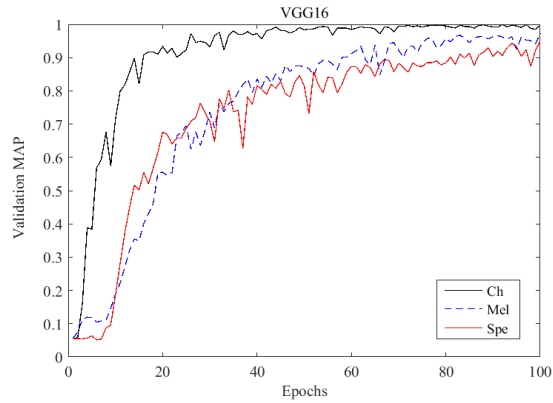

Fig.5 Variation of validation MAP with epochs increasing(VGG16)

For the TF model, the MAP arrives in the highest place faster than the VGG16 model. It means that the TF model is more efficient. Similar to the VGG16 model, the TF-Ch model performs the best, but the gaps of MAP between different spectrograms become smaller. Also, their training efficiencies are almost the same.

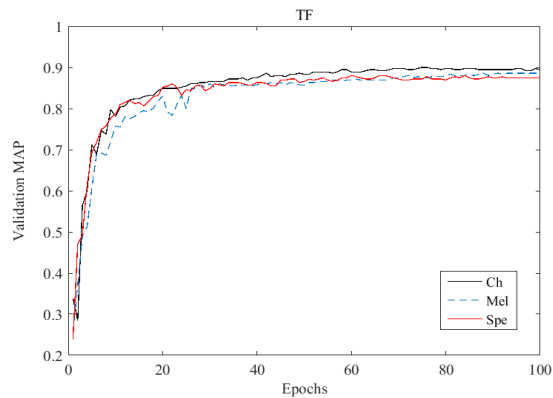

Fig.6 Variation of validation MAPs with epochs increasing(TF)

A comparison between validation MAPs and test MAPs was shown in Figure.7. Compared to the VGG16 models, the MAPs of the TF models are generally lower. The highest relative error was 9.92% when the Ch spectrogram sample set was fed. For the TF model, the parameters of feature extraction part were frozen, and they were not optimized during training. So it is reasonable that the MAPs are lower on validation sets and test sets. In later discussions, we will demonstrate that the multi-channel model can overcome this disadvantage.

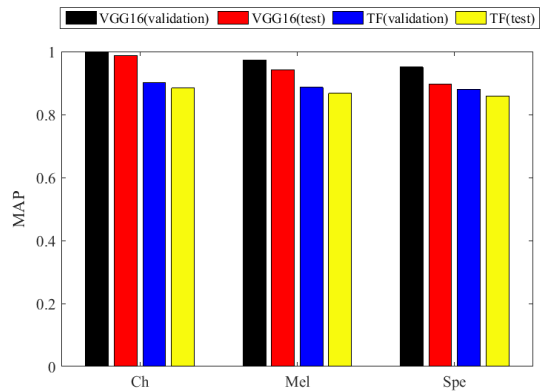

Fig.7 Validation and test MAP with different models and different inputs

Two fusion modes, including feature fusion and result fusion, were evaluated in this part. To investigate which fusion mode is better, three spectrogram sample sets, including Ch, Mel and Spe, were fed to two multi-channel models. Figure 8 shows the variation of validation MAPs of two multi-channel models. A comparison between validation MAPs and test MAPs was shown in Figure 9. The Re-fuse model gains the best MAP (0.9998) during epoch 8, whereas the Fe-fuse model gains a slightly lower MAP (0.9373) during the 85th epoch.

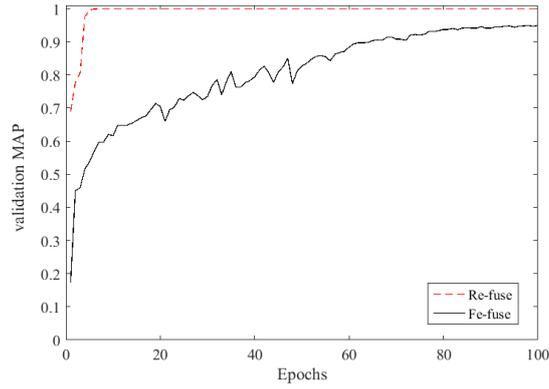
Fig.8 The variation of MAPs with increasing epochs (multi-channel models)

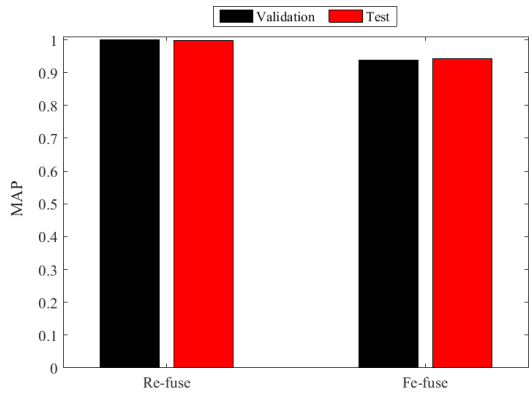
Fig.9 Validation and test MAP with different multi-channel models

### 3.3 Different models with different durations of spectrogram

Different durations of spectrogram may affect the performance of identification models. We chose the durations of 100ms, 300ms and 500ms and fed different sample sets to TF models and multi-channel models. A comparison of test MAPs between different models was shown in Figure 10. We found that when the spectrogram sample sets of the 300ms duration were fed in all the models, the performances are the best, and the test MAPs are the highest. The worst performances come from the 100ms duration.

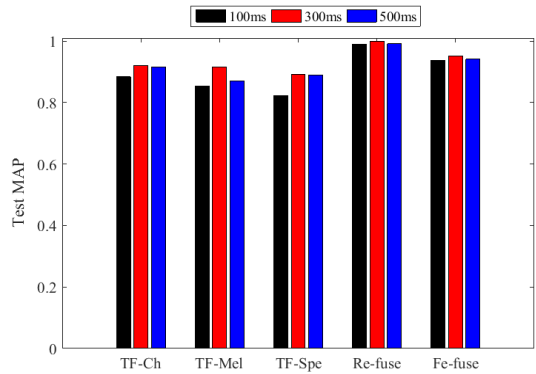
Fig 10 Comparison of test MAP between different models and different durations

## 4 Discussions

Prior work has documented the advantages of automatic bird species identification from vocalizations. Deep learning has been introduced to this field. They focused mainly on improving the accuracy of identification by building deeper or bigger models but ignored the demand of sample size. Practically, samples of certain bird species are often difficult to be collected, so that the sample size is too small to train a deep learning model, especially for some valuable and rare bird species

In this study, we used deep learning techniques and three kinds of time-frequency transforms to calculate the spectrograms, which were used to train three identification models based on parameter-based transfer learning. By comparison, we found that the Ch spectrograms with 300ms duration are most suitable for bird species identification based on vocalizations. Our results are in general agreement with H. Glotin's work(Glotin et al., 2016). Figure 11 lists the typical spectrograms of four bird species with 500ms duration. It is clear that the vocalization zones in Ch spectrograms are more concentrative and obvious than other kinds of spectrograms, which means that the Ch spectrogram is more favorable to the identification. It is believed that the syllable durations of the eighteen bird species are between 100ms and 250ms. When 100ms was chosen as the duration, a part of syllable may be cut off, so that the complete feature cannot be obtained. For 500ms, the number of samples is decreased, so that the performance of identification is negatively impacted. The appropriate duration should be selected according to the vocalization characteristics of the identified bird species. The adaptive duration should be designed in the future.

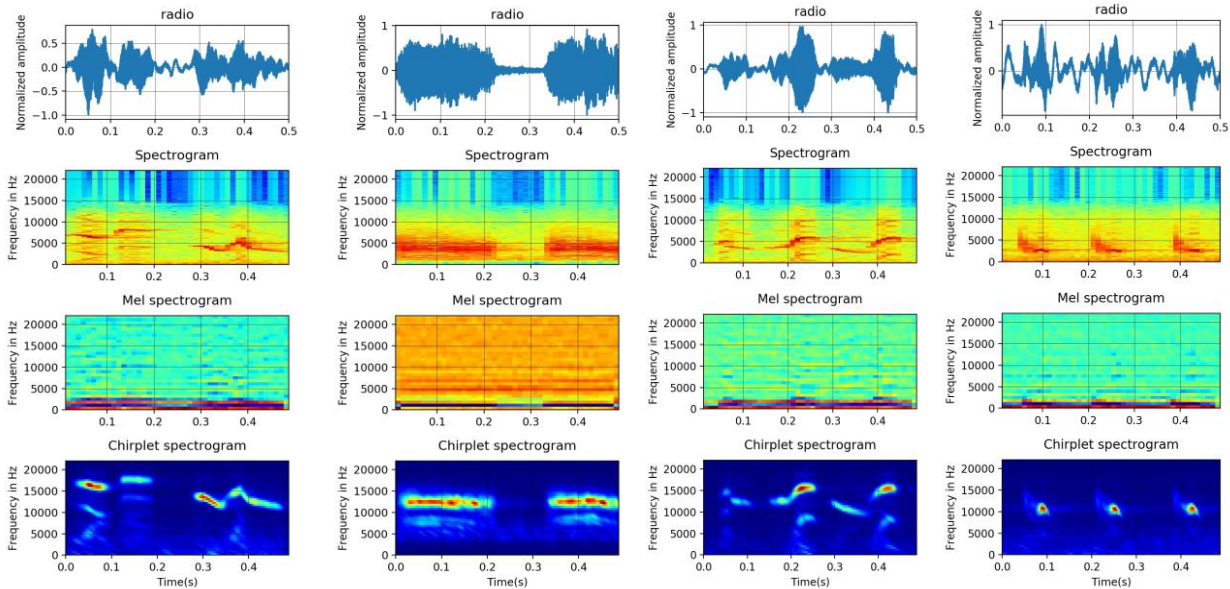

Fig.11 Typical spectrograms (from left to right, *P. venustulus*, *P. montanus*, *Phoenicurus auroreus* and *Sitta europaea*)

Because the MAPs of the TF models are lower than those of the VGG16 models, two kinds of multi-channel models were proposed. The Re-fusion model is better than the VGG16 model on both the identification efficiency and the MAP score. Furthermore, the number of trained parameters drops from 138 million (VGG16) to 13110. Therefore, the demand of sample is decreased. The multi-channel identification model can be effective to realize the bird species identification with a small sample size and a high MAP. Also, for the preprocessing of vocalization signals, there is no denoising operation, which means the proposed model has a certain function of anti-noise. The ability of anti-noise will be studied further.

## 6 Conclusions

Based on image feature of bird vocalization spectrogram, the method using deep learning model to identify the bird species has been proven to be an effective bird automated identification method. However, huge samples are needed to train deep learning model, and it is not appropriate for the identification of birds which are hard to collect samples. We introduced transfer learning to overcome this drawback. Most parameters are extracted from a pretrained model, only the parameters of the classifier are trained, so that the size demand of samples can be declined. Nevertheless, we found that the MAP scores of transfer learning models were lower than the original VGG16 model, as the maximum relative error was 9.92%. Meanwhile, the choice of spectrogram type made difference to identification performance. Although Mel spectrogram have been utilized in most previous bird species identification methods, the Ch spectrogram outperforms dramatically the Mel spectrogram in all our tests. With different durations of spectrograms as inputs, the performances are various. We recommend that researchers should choose suitable duration based on the vocalization features of birds to be recognized. At last, to raise the MAP of the transfer learning model, we proposed two fusion modes to form the multi-channel mode. The result about the fusion multi-channel modes has gained excellent performance on bird species identification, and the number of trained parameters is smaller than VGG16, which leads to small demand of samples. This is of great significance to the identification of birds with small samples or difficult to be recorded.

Acknowledgements

The work is supported by "the Fundamental Research Funds for the Central Universities (NO. 2017JC14)"